# Performance Analysis of Faults Detection in Wind Turbine Generator Based on High-Resolution Frequency Estimation Methods

CHAKKOR SAAD
Department of Physics,
Team: Communication and
Detection Systems, University of
Abdelmalek Essaâdi, Faculty of
Sciences, Tetouan, Morocco

BAGHOURI MOSTAFA
Department of Physics,
Team: Communication and
Detection Systems, University of
Abdelmalek Essaâdi, Faculty of
Sciences, Tetouan, Morocco

HAJRAOUI ABDERRAHMANE
Department of Physics,
Team: Communication and
Detection Systems, University of
Abdelmalek Essaâdi, Faculty of
Sciences, Tetouan, Morocco

*Abstract*—Electrical energy production based on wind power has become the most popular renewable resources in the recent years because it gets reliable clean energy with minimum cost. The major challenge for wind turbines is the electrical and the mechanical failures which can occur at any time causing prospective breakdowns and damages and therefore it leads to machine downtimes and to energy production loss. To circumvent this problem, several tools and techniques have been developed and used to enhance fault detection and diagnosis to be found in the stator current signature for wind turbines generators. Among these methods, parametric or super-resolution frequency estimation methods, which provides typical spectrum estimation, can be useful for this purpose. Facing on the plurality of these algorithms, a comparative performance analysis is made to evaluate robustness based on different metrics: accuracy, dispersion, computation cost, perturbations and faults severity. Finally, simulation results in MATLAB with most occurring faults indicate that ESPRIT and R-MUSIC algorithms have high capability of correctly identifying the frequencies of fault characteristic components, a performance ranking had been carried out to demonstrate the efficiency of the studied methods in faults detecting.

## I. INTRODUCTION

The increasing demand in energy over the world, as well as the growth in the prices of the energy fossil fuels resources and it exhaustion reserves in the long run, furthermore the commitment of the governments to reduce greenhouse gases emissions have favored the research of others energy sources. In this context, the recourse to renewable energy becomes a societal choice. The development of this alternative is encouraged because it offers natural, economic, clean and safe resources. Among the renewable energies, wind energy which has been progressed in a remarkable way in these recent years. It provides a considerable electrical energy production with fewer expenses with exception of construction and maintenance budget. Actually, wind energy investment has increased by multiplication of the wind parks capacities. This contributes greatly to the expansion of terrestrial and offshore wind parks. These parks are usually installed in far locations, difficult to access, subject to extreme environmental conditions. Therefore, a predictive monitoring scheme of wind turbines, allowing an early detection of electromechanical faults, becomes essential to reduce maintenance costs and ensure continuity of production. It means that stopping a wind installation for unexpected failures could lead to expensive repair and to lost production. This operating stopping becomes critical and causes very significant losses. For these reasons, there is an increase need to implement a robust efficient maintenance strategy to ensure uninterrupted power in the modern wind systems preventing major component failures, facilitating a proactive response, minimizing downtime and maximizing productivity [1], [8]. To anticipate the final shutdown of wind generators, on-line condition monitoring would be the most efficient technique because it allows the assessment of the health status of an operating machine by analysis of measured signals continuously [8]. Different types of sensors can be used to measure physical signals to detect the faults with various existing methods [4], [7], [9], [10], [12].

This is why reliability of wind turbines becomes an important topic in scientific research and in industry.

Most of the recent researches have been oriented toward electrical monitoring, with focus on the generator stator current. One of the most popular methods for fault diagnosis is the current signature analysis (CSA) as it is more practical and less costly [3], [4], [9], [10], [12]. Within the last decade many studies based on signal processing techniques have been conducted to detect electric machine faults prior to possible catastrophic failure. These researches initially developed for electric motor can be easily adapted to wind turbine generator. Furthermore, with recent digital signal processor (DSP) technology developments, motor and generator fault diagnosis can now be done in real-time [3]. Among signal processing techniques, non-parametric, parametric and high resolution or subspace methods (HRM) are widely adopted in machine diagnosis. They can be used for spectral estimation [10], [15], [16], [23], [24]. However,

This research work carried out in this direction with subspaces methods not highlight metrics of accuracy, robustness level of each approach related to the failures severity and computation time which is a key parameter in the context of a real-time integration. Otherwise, an investigation focused on the mean square error (MSE) and on the variance of faults harmonic detection must be done to evaluate the accuracy and detection robustness especially when the parameters of the signal, containing the faults in formations, will changes according to constraints of the application [18]. The main object of this study is to search a robust high resolution detection method for condition supervision, suitably adapted for implementation in wind generator.

## II. RELATED WORK

In the literature review, many research studies applying enhanced signal processing techniques and advanced tools have been commonly used in the wind generator stator current to monitor and to diagnose prospective mechanical or electrical faults. As known, these faults cause a modulation impact in the magnetic field of the wind generator, which is reflected by the appearance of a significant harmonics (peaks) in the stator current spectrum [8]. Nevertheless, these techniques are inappropriate because they have drawbacks such as high complexity, poor resolution and/or may suffer from some limitations. However, some failures are characterized by non-stationary behaviors [8], [14]. For this reason some researchers are leaning particularly toward methods adapted for non-stationary signals, such as time-frequency analysis, spectrogram, the wavelet decomposition (scalogram), Wigner-Ville representation, Concordia Transform (CT) and the Hilbert-Huang transform [13], [31].

In the first hand, in [7] a statistical diagnosis approach is proposed based on residues analysis of the electrical machine state variables by the use of the Principal Components Analysis method (PCA) for faults detection in Offshore Wind Turbine Generator. The aim drawback of this approach is that the detection efficiency requires a good choice of the principal components number. Some researchers are proposed failures diagnosis of wind turbines generators using impedance spectroscopy (IS) [27]. On the other hand, the periodogram and its extensions which are evaluated through a Fast Fourier Transform (FFT) is not a consistent estimator of the PSD because its variance does not tend to zero as the data length tends to infinity. Despite of this drawback, the periodogram has been used extensively for failure detection in research works [12], [17]. The (FFT) does not give any information on the time at which a frequency component occurs. Therefore, the Short Time Fourier Transform approach (STFT) is used to remove this shortcoming. A disadvantage of this approach is the increased sampling time for a good frequency resolution [32]. The discrimination of the frequency components contained within the signal, is limited by the length of the window relative to the duration of the signal [25]. To overcome this problem, in [8] and in [13] Discrete Wavelet Transform (DWT) is used to diagnose failures under transient conditions for wind energy conversion systems by analyzing frequencies with different resolutions. This method facilitates signal interpretation because it operates with all information contained in the signal by time-frequency redistribution.

One limitation of this technique that its gives a good time resolution and poor frequency resolution at high frequencies, and it provides a good frequency resolution and poor time resolution at low frequencies [12], [28]. Due mainly to their advantages, in [26] parametric methods have improved performance though they are affected by an adequate signal to noise ratio (SNR) level. High resolution methods (HRM) can detect frequencies with low SNR. They have been recently introduced in the area of induction motors and wind generators faults diagnosis by the application of multiple signal classification (MUSIC) method [2], [6], [29]. MUSIC and its zooming methods are conjugated to improve the detection by identifying a large number of frequencies in a given bandwidth [2], [28], [30].

Moreover, eigen analysis methods are especially suitable in case that the signal components are sinusoids corrupted by additive white noise. These algorithms are based on an eigen decomposition of the correlation matrix of the noise corrupted signal. Another approach used is ESPRIT [20], [26] [33], [34], [35]. It allows and performs well determination of the harmonic parameters components with high accuracy. In fact, this paper investigates the most efficient high-resolution techniques to detect faults in wind turbine generator.

## III. FAULTS IN WIND TURBINE GENERATOR

The wind generator is subjected to various electro-mechanical failures that affect mainly five components: the stator, the rotor, the bearings, gearbox and/or air gap (eccentricity) [5]. These faults require a predictive detection to avoid any side effect causing a breakdown or a fatal damage. However, a recent literature surveys [36], [37] shows that these defaults require periodic monitoring to avoid any unforeseen deterioration. Recent researches have been directed toward stator current supervision. Particularly, the current spectrum is analyzed to extract the frequency components introduced by the fault. A summary of wind turbines faults and theirs related frequencies are presented in Table I.

TABLE I. WIND TURBINES FAULTS SIGNATURES

| Failure | Harmonic Frequencies | Parameters |
|---|---|---|
| Broken rotor bars | $f_{brb} = f_0 \left[ k \left( \dfrac{1-s}{p} \right) \pm s \right]$ | $k = 1, 3, 5, ...$ |
| Bearing damage | $f_{bng} = \left| f_0 \pm k f_{i,o} \right|$ | $k = 1, 3, 5, ...$ $f_{i,o} = \begin{cases} 0.4\, n_b\, f_r \\ 0.6\, n_b\, f_r \end{cases}$ |
| Misalignment | $f_{mis} = \left| f_0 \pm k f_r \right|$ | $k = 1, 3, 5, ...$ |
| Air gap eccentricity | $f_{ecc} = f_0 \left[ 1 \pm m \left( \dfrac{1-s}{p} \right) \right]$ | $m = 1, 2, 3, ...$ |

Where $f_0$ is the electrical supply frequency, $s$ is the per-unit slip, $p$ is the number of poles, $f_r$ is the rotor frequency, $n_b$ is the bearing balls number, $f_{i,o}$ is the inner and the outer frequencies depending on the bearing characteristics, and $m, k \in \mathbb{N}$ [8], [12], [26].

## IV. Wind Generator Stator Current Model

To study the mentioned faults detection methods, the current will be denoted by the discrete signal x[n], which is obtained by sampling the continuous time current every $T_s=1/F_s$ seconds. The induction machine stator current x[n] in presence of mechanical and/or electrical faults can be expressed as follows [26]:

$$x[n] = \sum_{k=-L}^{L} a_k \cos\left(2\pi f_k(\omega(n)) \times \left(\frac{n}{F_s}\right) + \varphi_k\right) + b[n] \quad (1)$$

Where x[n] corresponds to the $n^{th}$ stator current sample, b[n] is a gaussian noise with zero mean and a variance equals to $\sigma^2 = 10^{-4}$ i.e. $b[n] \sim (0, 10^{-4})$. L is the number of sidebands introduced by the fault. The parameters $f_k(\omega)$, $a_k$, $\varphi_k$ correspond to the frequency, the amplitude and the phase of the $k^{th}$ component, respectively. $\omega(n)$ is a set of parameters to be estimated at each time n depending on the studied fault. The time and space of harmonics are not considered in this paper. The problem to solve is treated as a statistical estimation problem. It is an estimation of the fundamental frequency, the characteristic faults frequencies, and their amplitudes by the computation of the current spectrum from the stator current samples x(n).

## V. High-Resolution Frequency Estimation Methods

In this section, a brief description of each studied high resolution method and its main features are presented. The subspace frequency estimation methods rely on the property that the noise subspace eigenvectors of a Toeplitz autocorrelation matrix are orthogonal to the eigenvectors spanning the signal space. The model of the signal in this case is a sum of random sinusoids in the background of noise of a known covariance function. Among these methods, Prony method which is used for modeling sampled data as a linear combination of exponential functions. Although, it allows extracting P sinusoid or exponential signals from time data series, by solving a set of linear equations [15], [16], [24]. The signal s(n) is assumed equal to a sum of damped sines verifies the following recursive equation:

$$s(n) + b_1 s(n-1) + \cdots + b_{2P} s(n-2P) = 0 \quad (2)$$

$$B(z) = z^{2P} + b_1 z^{2P-1} + \cdots + b_{2P} \quad (3)$$

Polynomial (3) has 2P complex conjugate roots given by:

$$z_k = \rho_k e^{\pm 2\pi j f_k} \quad (4)$$

It's possible to calculate $b_k$, then the roots $z_k$ and therefore the frequencies $f_k$ and the damping coefficients $\rho_k$.

Unlike the methods using the periodogram, even with windowing, the high resolution methods are such that the error tends to zero when SNR→ ∞.

The Pisarenko Harmonic Decomposition PHD relies on eigendecomposition of correlation matrix which is decomposed into signal and noise subspaces. This method is the base of advanced frequency estimation methods. It has a limited practical use due to its sensitivity to noise [24], [15], [19], [22], [24], [33]. The eigenvector v associated with the smallest eigenvalue of the (2P+1) order covariance matrix $R_x$ of the observation has, as its components, the coefficients of the recursive equation (2) associated with the frequencies of the signal s(n). Then, the 2P degree polynomial B(z) is constructed based on v [19], [39]. The 2P complex conjugate roots $z_k$ are extracted from it, which leads us to the frequencies:

$$f_k = \frac{1}{2\pi}\arg(z_k) \quad (5)$$

For MUSIC (Multiple Signal Classification) approach, it is the improved version of Pisarenko method where M-dimensional space is split into signal and noise subspaces using many noise eigenfilters. The size of time window is taken to be M > P+1. Therefore, the dimension of noise subspace is greater than one and is equal to M-P. Averaging over noise subspace gives improved frequency estimation. Once the eigendecomposition of correlation matrix is calculated, it's used to find the (M x (M-P)) matrix G constructed from the (M-P) eigenvectors associated with the (M-P) smallest eigenvalues. Afterwards, the (M x M) matrix $GG^H$ is calculated to find the coefficients of the polynomial equation [6], [21], [22], [23], [24], [29], [33], [38]:

$$\tilde{Q}(z) = \begin{bmatrix} z^{M-1} & \cdots & z & 1 \end{bmatrix} GG \begin{bmatrix} 1 & z & \cdots & z^{M-1} \end{bmatrix}^T \quad (6)$$

Then, the estimation of the P frequencies values can be achieved as following:

$$f_k = \frac{\theta_k}{2\pi} \quad (7)$$

Two possibilities are available:

*1)* Calculating the 2(M-1) roots of $\tilde{Q}(z)$, then keeping the P stable roots that are closest to the unit circle. This is called the Root-MUSIC method.

*2)* Finding the P minima of $\tilde{Q}(e^{j\theta_k})$, using FFT function. This is called the FFT-MUSIC method.

Another method is Eigenvector (EV), this technique estimates the exponential frequencies from the peaks of eigenspectrum as follows:

$$\hat{P}_{EV} = \frac{1}{\sum_{i=P+1}^{M} \frac{1}{\lambda_i}\left|e^H v_i\right|^2} \quad (8)$$

However, with estimated autocorrelations, the EV method differs from MUSIC and produces fewer spurious peaks.

The last method is ESPRIT (Estimation of Signal Parameter via Rotational Invariance Technique) algorithm which allows determining and detecting the parameters of harmonic components with very high accuracy both in frequency and in amplitude estimation independently of the window length. Furthermore, it's a suitable approach to providing reliable spectral results without synchronization effects [20], [21], [22], [23], [24], [33], [38]. It is based on naturally existing shift invariance between the discrete time series which leads to rotational invariance between the corresponding signal subspaces. The eigenvectors U of the autocorrelation matrix of the signal define two subspaces (signal and noise subspaces) by using two selector matrices $\Gamma_1$ and $\Gamma_2$.

$$S_1 = \Gamma_1 U, \quad S_2 = \Gamma_2 U \quad (9)$$

The rotational invariance between both subspaces leads to the following equation:

$$S_1 = \Phi S_2 \quad (10)$$

Where:

$$\Phi = \begin{bmatrix} e^{j2\pi f_1} & 0 & \cdots & 0 \\ 0 & e^{j2\pi f_2} & \cdots & 0 \\ \vdots & \vdots & \ddots & \vdots \\ 0 & 0 & \cdots & e^{j2\pi f_M} \end{bmatrix} \quad (11)$$

The matrix $\Phi$ contains all information about M components frequencies, and the estimated matrices S can contain errors. Moreover, the TLS (total least-squares) approach finds the matrix $\Phi$ as minimization of the Frobenius norm of the error matrix.

Another interesting eigendecomposition method is the minimum norm (MN) algorithm. Instead of forming an eigenspectrum that uses all of the noise eigenvectors. It uses a single vector which is constrained to lie in the noise subspace, and the complex exponential frequencies are estimated from the peaks of the frequency estimation function given by:

$$\hat{P}_{MN}\left(e^{j\omega}\right) = \frac{1}{\left|e^H a\right|^2}, \quad a = \lambda P_n u_1 \quad (12)$$

The problem, therefore, is to determine which vector in the noise subspace minimizes the effects of the spurious zeros on the peaks of $\hat{P}_{MN}\left(e^{j\omega}\right)$.

It is proposed in this work to apply these methods for detection of different wind turbine generator faults.

## VI. COMPARATIVE PERFORMANCE ANALYSIS

The performance (error of estimation) of the subspace methods has been extensively investigated in the literature, especially in the context of the Direction of Arrival (DOA) estimation [33]. In this section, to evaluate the efficiency of the above mentioned fault detectors, with respect to the computation speed, accuracy, degree of frequency estimation dispersion for different level of SNR with a fixed values of fault amplitude.

The faults severity detection is also studied by varying the faults amplitude $a_{-1}$, $a_1$ in the interval $[0, 0.2a_0]$. The previous frequency estimation methods are applied under different scenarios by simulation in Matlab for a faulty wind turbine generator using 2 pair poles, 4kW/50Hz, 230/400V. The induction generator stator current, showed in figure 1 for a window time of 0.25 s, is simulated by using the signal model described in (1) for the different failure cases described in table I. The parameters of the simulation are illustrated in table II and in table III.

With $f_k(\omega) = f_k(f_0, s, p, k, m)$ is a set of parameters to be estimated at each time n depending on the faults studied cases. Choosing between estimators is a difficult task. Therefore, some quality criteria are needed to determine the best one. The comparison of mean square error (MSE) defined by equation (13) would be helpful for theoretical assessment of accuracy for this purpose.

$$MSE = \frac{1}{N}\sum_{i=1}^{N}\left(\hat{f}_i - f_i\right)^2 \quad (13)$$

$\hat{f}_i$ is the estimated fault frequency

$f_i$ is the exact fault frequency

N is the iterations number

TABLE II. PARAMETERS USED IN SIMULATIONS

| Parameter | Value |
|---|---|
| s | 0,033 |
| L | 2 |
| p | 2 |
| m | {1,2} |
| $f_0$ | 50 Hz |
| $f_r$ | 29,01 Hz |
| $n_b$ | 12 |
| k | {1,3} |
| n | 1600 |
| $F_s$ | 1000 Hz |
| iterations | 200 |
| SNR | [0,100] |
| Stator Current Amplitude $a_0$ | 10 A |
| Computation Processor | Intel Core2 Duo T6570 2,1 GHz |

TABLE III. FAULTS SIMULATION SCENARIOS

| | Frequencies (Hz) | | Amplitudes (A) | | Phase (rad) | |
|---|---|---|---|---|---|---|
| Fault | $\phi_{-1}$ | $\phi_1$ | $a_{-1}$ | $a_1$ | $\varphi_{-1}$ | $\varphi_1$ |
| Broken rotor bars | 22,53 | 70,83 | 1 | 1 | 0 | 0 |
| Inner Bearing damage | 89,25 | 367,74 | 1 | 1 | 0 | 0 |
| Misalignment | 79,01 | 137,03 | 1 | 1 | 0 | 0 |
| Air gap eccentricity | 74,18 | 98,35 | 1 | 1 | 0 | 0 |

Therefore, for each scenario the fault harmonic amplitude is fixed to $0.1a_0$ as shown in table III. Simulation results for the broken rotor bars fault frequency estimation shows in the figures 2, 3 the evolution of the MSE and the variance average depending on the variation of the SNR. It is very clear that for a stator current having a high level of noise in [0,30]dB, R-MUSIC and ESPRIT gives a nearly identical poor accuracy because these approaches have a high MSE and a high variance values almost constants due to theirs sensibility to noise. Afterwards, Prony and Pisarenko are almost identical and they present a medium accuracy value over other methods, hereafter come Min-Norm then EV method with a constant value of MSE and an increased variation of the variance average.

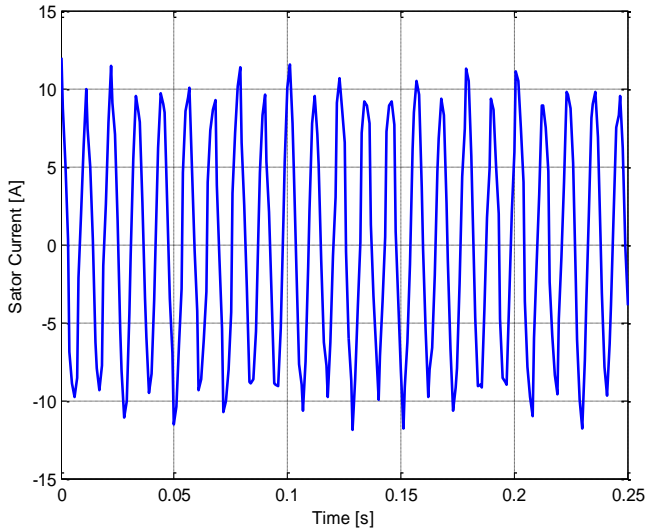

Fig. 1. Faulty induction generator stator current with noise (SNR=30 dB)

Moreover, it is noted that EV gives a good accuracy compared to Min-Norm due to it resistance to noise. Contrariwise, R-MUSIC then ESPRIT becomes more accurate when the SNR increases in [35,100]dB, the accuracy level of these methods exceeds that gives Prony and Pisarenko which have a medium value.

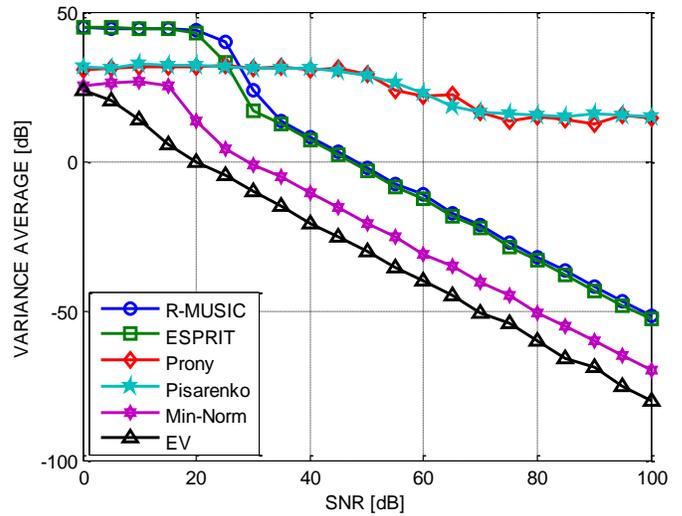

Fig. 3. Variance Average of Broken rotor bars fault frequency estimation

Whereas, figures 4 provide a statistical computation time description, it can be divided into three categories: fast computation methods for Min-Norm, Prony and Pisarenko followed by a medium speed computation for R-MUSIC then EV and ESPRIT come with a high computation time cost. This variation in computation speed for each method can be justified by the autocorrelation and the covariance matrix calculation in addition to the simulated stator current samples size and the sampling frequency used.

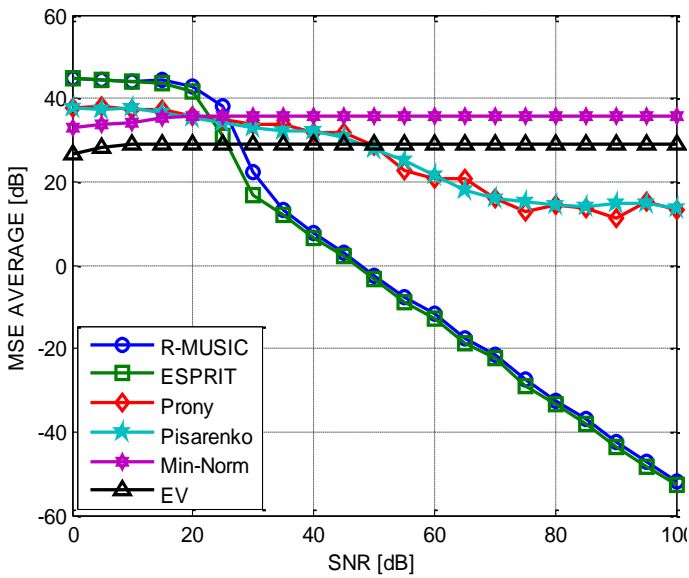

Fig. 2. Mean Square Error Average of Broken rotor bars fault frequency estimation

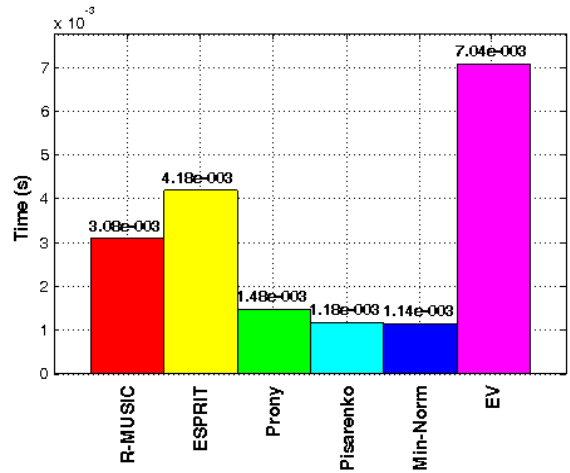

Fig. 4. Computational time cost of average Broken rotor bars fault detection for differents HRM

For inner bearing damage fault frequency estimation, based on the simulation results illustrated in figures 5 and 6, it can be noted that the accuracy of the methods can be classified into three levels: the first level in which ESPRIT and R-MUSIC leading to almost the same very high accuracy even at low SNR because their variance and MSE which decreases rapidly with increasing SNR, in the high level Min-Norm and EV are founded followed by Pisarenko and Prony in the medium accuracy level giving almost constant MSE and variance values.

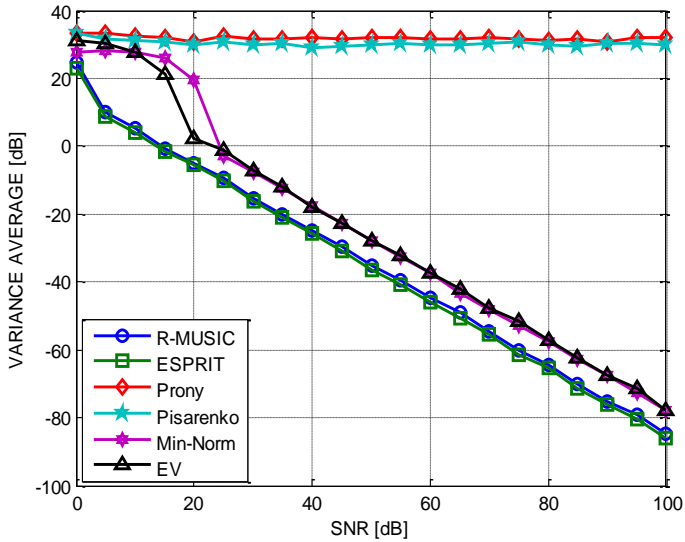

Fig. 5. Variance Average of Inner Bearing damage fault frequency estimation

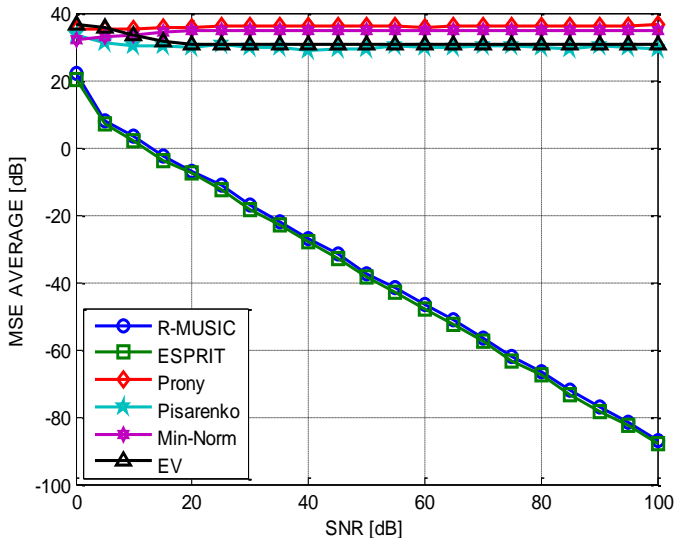

Fig. 6. Mean Square Error Average of Inner Bearing damage fault frequency estimation

This difference in accuracy is mainly due to more disparity in the faults frequency components values. Concerning the computation time, from figure 7, 10 and 13 it is noticed that the studied methods keep the same speed behavior observed in previous scenario.

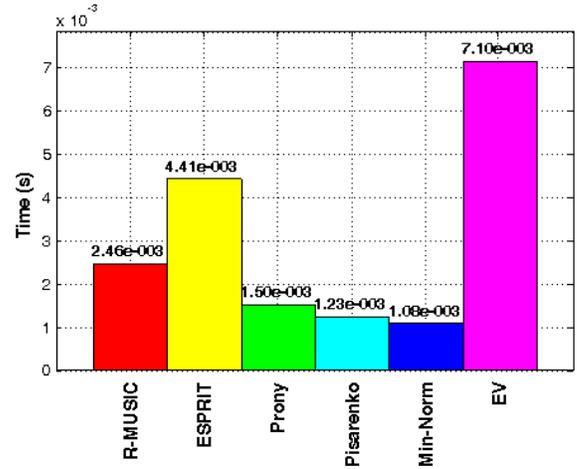

Fig. 7. Computational time cost average of Inner Bearing damage fault detection for differents HRM

Referred to figure 8 and 9, frequency detection precision for misalignment fault is very important and increases significantly by increasing SNR for ESPRIT and R-MUSIC, whereas this accuracy is modest for Prony and Pisarenko mostly beyond an SNR lower than 30dB, after this value EV, Prony and Pisarenko gives almost the same moderate exactitude except Min-Norm which presents a relatively good accuracy which does not reaches ESPRIT and R-MUSIC strictness.

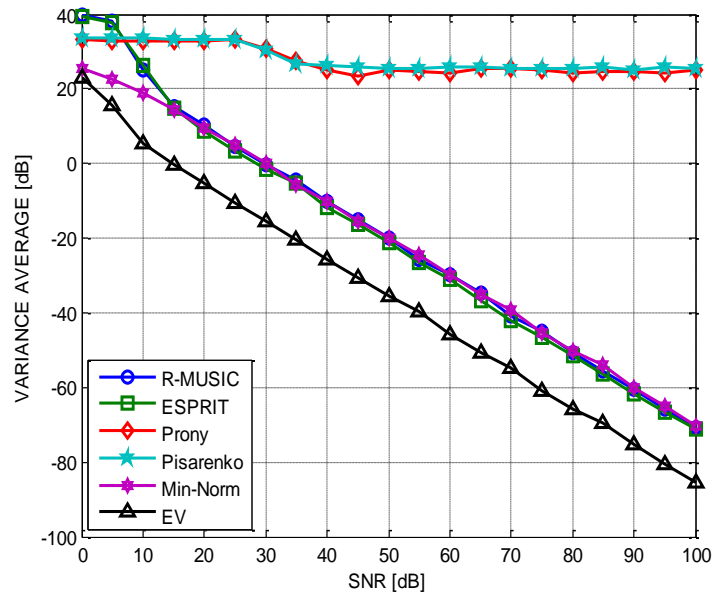

Fig. 8. Variance Average of Misalignment fault frequency estimation

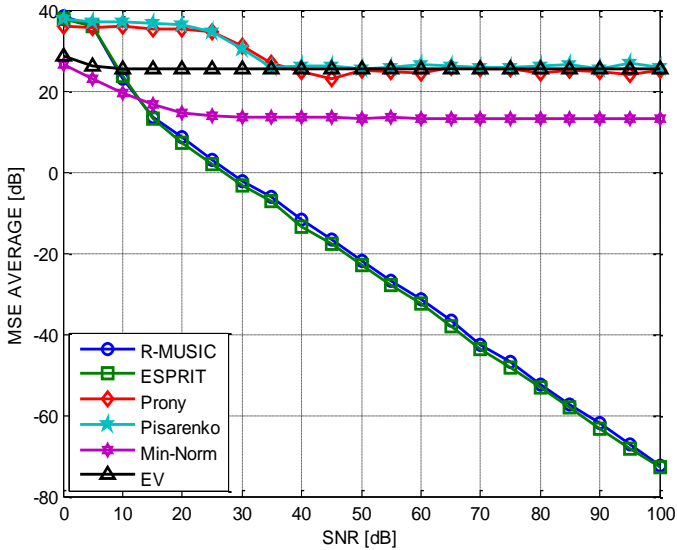

Fig. 9. Mean Square Error Average of Misalignement fault frequency estimation

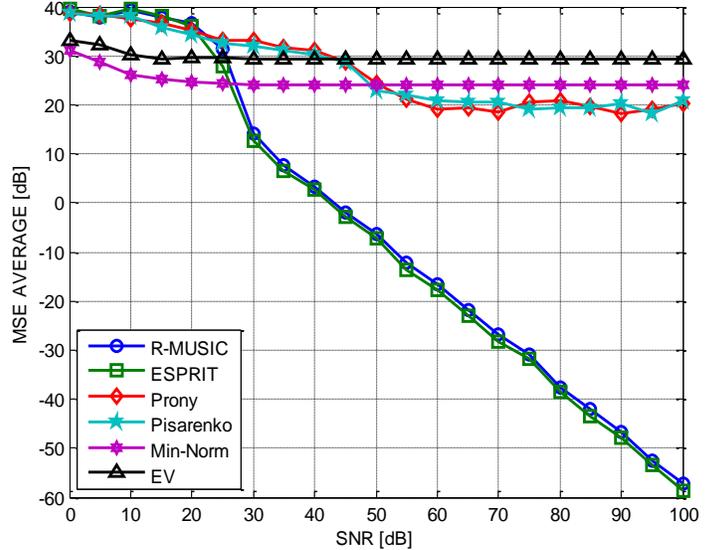

Fig. 11. Mean Square Error Average of Air gap eccentricity fault frequency estimation

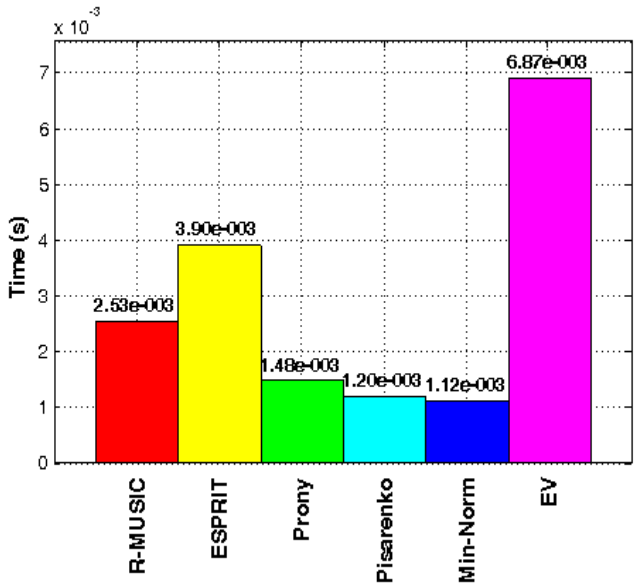

Fig. 10. Computational time cost average of Misalignement fault detection for differents HRM

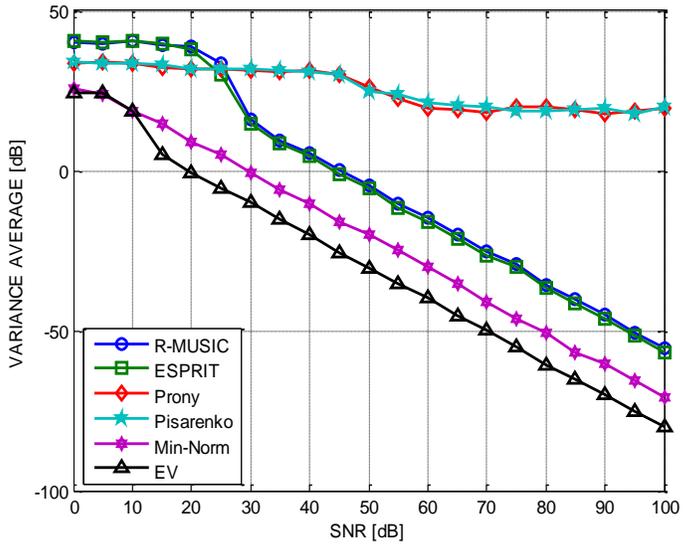

Fig. 12. Variance Average of Air gap eccentricity fault frequency estimation

It seems clearly in figures 11 and 12, for a noisy stator current generator in presence of an air gap eccentricity fault, that Min-Norm and EV are the best choice due to their good accuracy on the contrary of Prony, Pisarenko, R-MUSIC and ESPRIT which gives bad one, but when SNR exceeds 25dB it is observed that R-MUSIC and ESPRIT occupies the first place in accuracy followed by Prony and Pisarenko exceeding Min-Norm for an SNR above 50dB, from this point there are four degrees of quality estimation frequency:

the best one is that take ESPRIT and R-MUSIC, the second one is related to Prony and Pisarenko while others are modest.

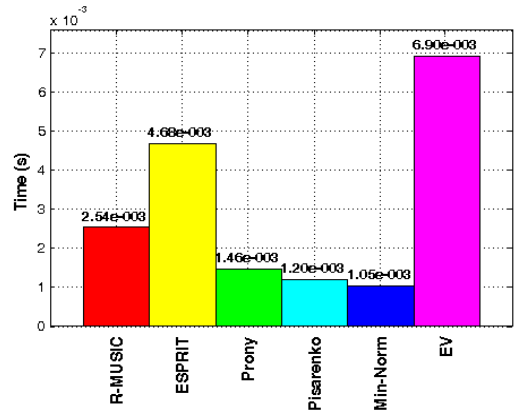

Fig. 13. Computational time cost average of Air gap eccentricity fault detection for differents HRM

To compare and to investigate the ability of the studied methods to detect and to identify clearly the fault frequency components for small amplitudes even in presence of an annoying noise, figures 14, 15, 16 and 17 shows the simulation results obtained for the different studied faults scenarios.

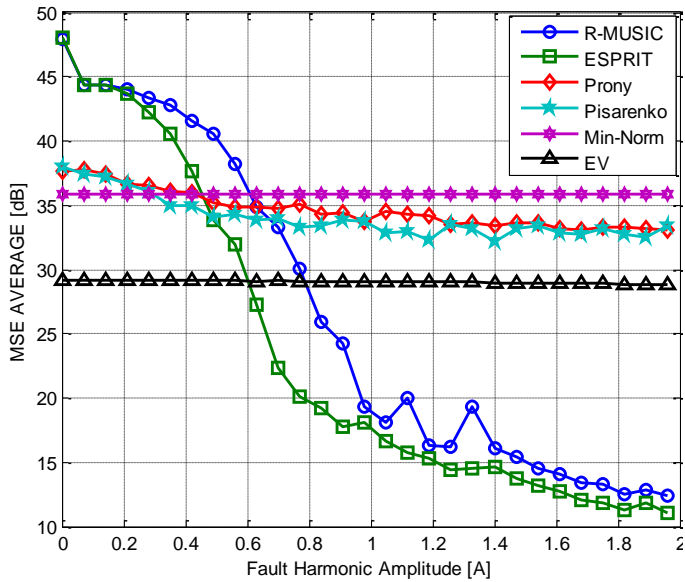

Fig. 14. Mean Square Error Average of Broken rotor bars fault frequency estimation depending on harmonic amplitude variation for differents HRM (SNR=30 dB)

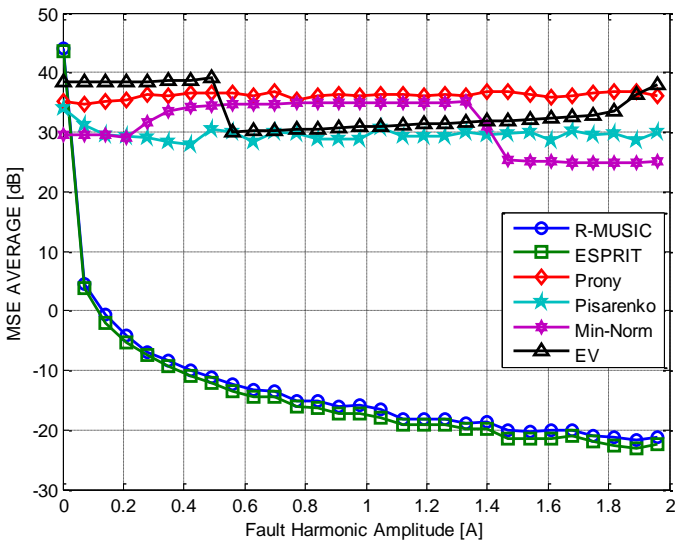

Fig. 15. Mean Square Error Average of Inner Bearing damage fault frequency estimation depending on harmonic amplitude variation for differents HRM (SNR=30 dB)

In a global vision, it is concluded that the ESPRIT and the R-MUSIC methods are very powerful in the detection of fault frequencies despite their corresponding amplitudes are very small, whereas EV and Min-Norm show some instabilities in this identification and their robustness remains limited, on the other hand Prony and Pisarenko have a poor degree of performance compared to ESPRIT and R-MUSIC.

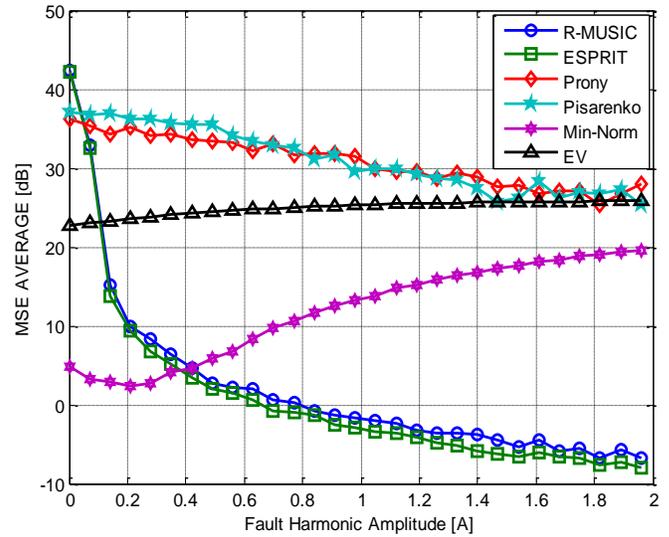

Fig. 16. Mean Square Error Average of Misalignement fault frequency estimation depending on harmonic amplitude variation for differents HRM (SNR=30 dB)

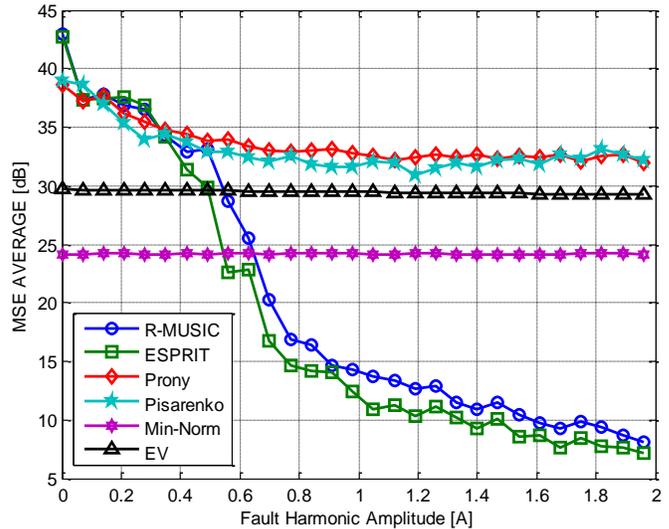

Fig. 17. Mean Square Error Average of Air gap eccentricity fault frequency estimation depending on harmonic amplitude variation for differents HRM (SNR=30 dB)

Generally, ESPRIT and R-MUSIC are even competitive. They have a good detection and resolution capabilities clearly outperform other studied methods. ESPRIT parametric spectral method, accurately and reliably showed high content of fault harmonics in the stator current, justifying their usefulness as a tool for spectral analysis of distorted electric signals in wind power generators although it high computation time cost. Prony and Pisarenko are not very much fruitful when noise increases. They have a limited practical use. EV and Min-Norm are good approaches but sometimes they give a risk of false estimates harmonics due to their roots of eigenvectors which does not correspond to the required frequency. As an outcome, these signal processing methods has been ordered according to the three evaluation criteria previously studied as shown in Table IV

TABLE IV.    PERFORMANCE CHARACTERISTICS COMPARAISON OF THE STUDIED PARAMETRIC SPECTRAL METHODS

| Method | Time | Accuracy | Risk | Rank |
|---|---|---|---|---|
| ESPRIT | medium | very high | none | 1 |
| R-MUSIC | medium | high | none | 2 |
| Min-Norm | small | medium | medium | 3 |
| EV | high | medium | medium | 4 |
| Pisarenko | small | low | medium | 5 |
| Prony | small | low | medium | 6 |

## VII. CONCLUSION

In this paper, it has been shown that the high-resolution spectrum estimation methods could be effectively used for wind turbine faults detection which can be achieved by on-line monitoring stator current spectral components produced by the magnetic field anomaly. These techniques aim to separate the observation space in a signal subspace, containing only useful information improving the spectral resolution. An investigation under different conditions is realized to measure robustness and to found efficient tools for detection. The accuracy of the estimation depends on the signal perturbation, fault severity level, the sampling frequency and on the number of samples taken into the estimation process. The comparison has proved the superiority of ESPRIT algorithm than the others followed by R-MUSIC which allows in all cases very high detection accuracy. However, their computation is slightly more complex than the others approaches which can affect their use in real-time implementation. Despite this, ESPRIT can be exploited to design an intelligent embedded system for diagnosis of electromechanical problems in wind turbines generators. As future work, the enhancement of the accuracy and the computation time cost so much more form an important defiance.

REFERENCES

[1] M.C. Mallikarjune Gowda et al, "Improvement of the Performance of Wind Turbine Generator Using Condition Monitoring Techniques", Proceedings of 7th International Conference on Intelligent Systems and Control (ISCO 2013), IEEE 2012

[2] Don-Ha Hwang et al, "Robust Diagnosis Algorithm for Identifying Broken Rotor Bar Faults in Induction Motors", Journal of Electrical Engineering & Technology, JEET, Vol. 9, No. 1, January 2014

[3] Hamid A. Toliyat et al., "Electric Machines Modeling, Condition Monitoring, and Fault Diagnosis", CRC Press Taylor & Francis Group NW 2013, ISBN-13: 978-1-4200-0628-5

[4] M. L. Sin, W. L. Soong and N. Ertugrul, "On-Line Condition Monitoring and Fault Diagnosis – A Survey" Australian Universities Power Engineering Conference, New Zealand, 2003.

[5] Shawn Sheng and Jon Keller et al, "Gearbox Reliability Collaborative Update", NREL U.S. Department of Energy, http://www.nrel.gov/docs/fy14osti/60141.pdf

[6] Shahin Hedayati Kia et al, "A High-Resolution Frequency Estimation Method for Three-Phase Induction Machine Fault Detection", IEEE Transactions on Industrial Electronics, Vol. 54, No. 4, AUGUST 2007.

[7] Ouadie Bennouna et al, "Condition Monitoring & Fault Diagnosis System for Offshore Wind Turbines", https://zet10.ipee.pwr.wroc.pl/record/425/files/invited_paper_3.pdf

[8] Elie Al-Ahmar et al, "Wind Energy Conversion Systems Fault Diagnosis Using Wavelet Analysis", International Review of Electrical Engineering 3, 4 (2008) 646-652, http://hal.univ-brest.fr/docs/00/52/65/07/PDF/IREE_2008_AL-AHMAR.pdf

[9] M.L. Sin et al, "Induction Machine On-Line Condition Monitoring and Fault Diagnosis – A Survey", http://www.academia.edu/416441/Induction_Machine_on_Line_Condition_Monitoring_and_Fault_Diagnosis_A_Survey

[10] K. K. Pandey et al, "Review on Fault Diagnosis in Three-Phase Induction Motor", MEDHA – 2012, Proceedings published by International Journal of Computer Applications® (IJCA)

[11] Niaoqing Hu et al, "Early Fault Detection using A Novel Spectrum Enhancement Method for Motor Current Signature Analysis", 7th WSEAS Int. Conf. on Artificial Intelligence, Knowledge Engineering and Data Bases (AIKED'08), University of Cambridge, UK, Feb 20-22, 2008

[12] E. Al Ahmar et al, "Advanced Signal Processing Techniques for Fault Detection and Diagnosis in a Wind Turbine Induction Generator Drive Train: A Comparative Study", IEEE Energy Conversion Congress and Exposition (ECCE), 2010, Atlanta : États-Unis (2010)

[13] El Houssin El Bouchikhi et al, "A Comparative Study of Time-Frequency Representations for Fault Detection in Wind Turbine", IECON 2011 - 37th Annual Conference on IEEE Industrial Electronics Society

[14] Francisco José Vedreño Santos, "Diagnosis of Electric Induction Machines in Non-Stationary Regimes Working in Randomnly Changing Conditions", Thesis Report, Universitat Politècnica de València, November 2013

[15] M. Hayes, "Digital signal processing and modeling", Wiley, New York, NY, 1996.

[16] T.T. Georgiou, "Spectral Estimation via Selective Harmonic Amplification", IEEE Trans. on Automatic Control, 46(1): 29-42, January 2001.

[17] T.T. Georgiou, "Spectral analysis based on the state covariance: the maximum entropy spectrum and linear fractional parameterization," IEEE Trans. on Automatic Control, 47(11): 1811-1823, November 2002.

[18] P. Billingsley, "Probability and Measure", second edition, John Wiley and Sons, New York, 1986.

[19] V.F. Pisarenko, "The Retrieval of Harmonics from a Covariance Function", Geophysics J. Roy. Astron. Soc. 33 (1973), 347-366.

[20] R. Roy and T. Kailath, "ESPRIT - Estimation of Signal Parameters via Rotational Invariance Techniques", IEEE Transactions on Acoustics, Speech, and Signal Processing. ASSP-37 (1989), 984-995.

[21] R.O Schmidt, "A Signal Subspace Approach to Multiple Emitter Location and Spectral Estimation", Ph.D. thesis, Stanford University, Stanford, CA, 1981.

[22] Sophocles J. Orfanidis, "Optimum Signal Processing", MCGRAW-HILL publishing company, New York, ny, 2nd edition 2007.

[23] John L. Semmlow, "Biosignal and Biomedical Matlab-Based Applications", Marcel Dekker, Inc New York 2004.

[24] Gérard Blanchet and Maurice Charbit, "Digital Signal and Image Processing using Matlab", ISTE USA 2006.

[25] Yassine Amirat et al, "Wind Turbine Bearing Failure Detection Using Generator Stator Current Homopolar Component Ensemble Empirical Mode Decomposition", IECON 2012 - 38th Annual Conference on IEEE Industrial Electronics Society

[26] El Houssin El Bouchikhi et al, "A Parametric Spectral Estimator for Faults Detection in Induction Machines", Industrial Electronics Society, IECON 2013 - 39th Annual Conference of the IEEE

[27] Mohamed Becherif et al, "On Impedance Spectroscopy Contribution to Failure Diagnosis in Wind Turbine Generators", International Journal on Energy Conversion 1, 3 (2013) pages 147-153.

[28] Ioannis Tsoumas et al, "A comparative study of induction motor current signature analysis techniques for mechanical faults detection, SDEMPED 2005 - International Symposium on Diagnostics for Electric Machines", Power Electronics and Drives Vienna, Austria, 7-9 September 2005

[29] Young-Woo Youn et al, "MUSIC-based Diagnosis Algorithm for Identifying Broken Rotor Bar Faults in Induction Motors Using Flux Signal, Journal of Electrical Engineering & Technology", JEET, Vol. 8, No. 2, 2013, pages 288-294.

[30] El Houssin El Bouchikhi et al, "Induction Machine Fault Detection Enhancement Using a Stator Current High Resolution Spectrum", IECON 2012 - 38th Annual Conference on IEEE Industrial Electronics Society

[31] Yassine Amirat et al, "Wind Turbines Condition Monitoring and Fault Diagnosis Using Generator Current Amplitude Demodulation", IEEE International Energy Conference and Exhibition (EnergyCon), 2010

[32] Neelam Mehala et al, "Condition monitoring methods, failure identification and analysis for Induction machines", International Journal of Circuits, Systems and Signal Processing, Issue 1, Volume 3, 2009, pages 10-17

[33] Zbigniew Leonowicz, "Parametric methods for time-frequency analysis of electric signals", Politechnika Wrocławska Wroclaw University of Technology, Poland, 2nd edition January 2007.

[34] Przemyslaw Janik et al, "Advanced Signal Processing Methods for Evaluation of Harmonic Distortion Caused by DFIG Wind Generator", 16th PSCC, Glasgow, Scotland, July 14-18, 2008.

[35] H. C. So et al, "Linear Prediction Approach for Efficient Frequency Estimation of Multiple Real Sinusoids: Algorithms and Analyses", IEEE Transactions on Signal Processing, Vol. 53, No. 7, July 2005

[36] Tavner, P. "How Are We Going to Make Offshore Wind Farms More Reliable?", SUPERGEN Wind General Assembly on March 20, 2011 at Durham University, UK.

[37] Sheng, S. "Investigation of Various Wind Turbine Drive train Condition Monitoring Techniques", Wind Turbine Reliability Workshop, August 2−3, 2012 Albuquerque, NM.

[38] André Quinquis, "Digital Signal Processing using MATLAB", ISTE Ltd, London UK, 2008

[39] Łobos T. et al., "Advanced signal processing methods of harmonics and interharmonics estimation", IEE Seventh International Conference on Developments in Power System Protection, Amsterdam, 9–12 April 2001, pp. 315–318.